\documentclass[final]{svjour2}
\usepackage{graphicx}
\usepackage{rotating}
\usepackage{amssymb}
\usepackage{mathptmx}
\usepackage[numbers]{natbib}
\makeatletter
\journalname{Journal of Low Temperature Physics}

\bibpunct{}{}{,}{s}{}{,}

\begin{document}

\newcommand{\hdblarrow}{H\makebox[0.9ex][l]{$\downdownarrows$}-}
\title{Bose-Einstein condensation of magnons in superfluid $^3$He}

\author{Yu.M. Bunkov$^1$ and G.E. Volovik$^{2,3}$}

\institute{1: MCBT, Institute Neel, CNRS/UJF, Grenoble, 38042, France\\
\email{yuriy.bunkov@grenoble.cnrs.fr}
\\2: Low Temperature Laboratory, Helsinki University of
Technology, Finland\\
\email{volovik@boojum.hut.fi} \\3: L.D. Landau Institute for
Theoretical Physics,
 Moscow, Russia}

\date{XX.XX.2007}

\maketitle


\begin{abstract}

   The possibility of Bose-Einstein condensation of excitations has been discussed for a long time.  The phenomenon of the phase-coherent precession of
magnetization in superfluid $^3$He and the related effects of spin superfluidity  are based
on the true Bose-Einstein condensation of magnons. Several different states of coherent precession has been observed in $^3$He-B: homogeneously precessing domain (HPD); persistent signal formed by $Q$-balls at very low temperatures;  coherent precession with
fractional magnetization; and  two new modes of the coherent precession  in compressed aerogel. In compressed aerogel  the coherent precession  has been also found in $^3$He-A. Here we demonstrate that all  these cases are examples of a Bose-Einstein condensation of magnons, with the  magnon interaction term in the Gross-Pitaevskii equation being provided by different types of spin-orbit coupling
 in the background of the coherent precession.

\end{abstract}

\section{Introduction}

Bose-Einstein condensation (BEC) is a  phenomenon of  formation
of collective quantum state, in which the macroscopic number
of particles is governed by a single wave function. The phenomenon
of Bose-Einstein condensate was predicted by Einstein in 1925.  For
a review see, for example Ref. \cite{revBEC}. The almost perfect BEC
state was observed in ultra could atomic gases. In Bose liquids, the
BEC is strongly modified by interactions, but still remains the key
mechanism for the formation of a coherent quantum state in Bose systems, which
exhibits the phenomenon of superfluidity characterized by  non-dissipative superfluid
mass currents discovered first in $^4$He by P.L. Kapitza \cite{Kapitza}. Superfluidity proved to be a more general phenomenon: superfluid mass current has been found in Fermi liquid $^3$He;  superfluidity of electric charge -- superconductivity --  is known in metals; quantum chromodynamics is characterized by superfluidity of chiral charge; color superfluidity is discussed in quark matter and baryonic superfluidity -- in neutron stars; etc. Here we discuss the magnon BEC which leads to superfluidity of spin current.

Strictly speaking, the theory of superfluidity and Bose-Einstein
condensation is applicable to systems with conserved $U(1)$ charge  or
particle number. However, it can be extended to systems with a
weakly violated conservation law. This means that it can be
applicable to a system of sufficiently long-lived quasiparticles --
discrete quanta of energy that can be treated as condensed matter counterpart of
elementary  particles. In
magnetically ordered materials, the corresponding propagating excitations are magnons -- quanta of spin waves.
Under stationary conditions the density of thermal magnons is small, but they can be pumped  by resonance radio-frequency (RF) field (magnetic resonance).
One may expect that at very low temperatures, the non-equilibrium gas of magnons could  live a relatively long time, sufficient for formation of coherent magnon condensate.

Recently there appeared a number of articles, where authors
claimed the observation of BEC of quasiparticles: excitons
\cite{exit} and magnons \cite{Democ}.
According to Snoke \cite{Snoke} to claim the observation of
BEC one should demonstrate the spontaneous emergence of coherence.
And, even better, to show the interference between two condensates.
Since the spontaneous coherence has not
been observed directly, the claim of observation of BEC in the above article is still under question.

In superfluid $^3$He-B, the formation of a coherent state of  magnons was discovered about 20 years ago \cite{HPD}. In  pulsed NMR experiments  the spontaneous formation of domain with fully phase-coherent
BEC of magnons has been observed even in the presence of inhomogeneous magnetic field. This domain was called the Homogeneously Precessing Domain (HPD). The main feature of HPD is the induction decay signal, which rings in many orders of magnitude longer, than prescribed by inhomogeneity of magnetic field. This means that  spins precess NOT with a local Larmor frequency, but  precess coherently with a common frequency and phase. This BEC can be also created and
stabilized by continuous NMR pumping. In this case the NMR frequency
plays a role of magnon chemical potential, which determines the density of magnon condensate. The interference between two condensates has also been demonstrated.
 It was shown that HPD exhibits all the properties of spin superfluidity (see Reviews
\cite{FominLT19,BunkovHPDReview}). The main property is the existence of spin
supercurrent,  which transports the magnetization on a macroscopic
distance more than 1 cm long. This spin supercurrent  flows separately from the mass current, which is essentially different from the spin-polarized  $^3$He-A$_1$, where  spin is transported by the mass current.
Also the  related phenomena have been observed: spin
current Josephson effect; phase-slip processes at the critical
current; and spin current vortex  --  a topological defect which is
the analog of a quantized vortex in superfluids and of an Abrikosov
vortex in superconductors; etc.

The spin-orbit coupling, which is responsible for the interaction between
magnons, is relatively small in $^3$He-B. As a result  HPD
represents almost pure BEC of magnons \cite{V,B}. In typical $^3$He experiments, the critical temperature of magnon condensation is 3 orders of magnitude higher, than the temperature of superfluid transition; i.e.  magnons undergo the condensation as soon as chemical potential and spin-orbit coupling allow for this process.
The superfluid $^3$He is a very unique complex macroscopic quantum
system with broken spin, orbital and gauge symmetries, where the spin-orbit coupling  due to dipole-dipole interaction between the spins of $^3$He atoms can be varied experimentally. Under different conditions one can observe different types of the BEC of the gas
of magnons in $^3$He-B \cite{HPD2};
non-topological solitons called $Q$-balls in high energy  physics \cite{Qbal}; and also magnon BEC in $^3$He-A \cite{JapCQS}.

\section{Coherent precession as magnon BEC}

As distinct from the static equilibrium magnetic states with broken symmetry,  the phase-coherent  precession is the dynamical state which experiences the off-diagonal long-range order:
\begin{equation}
\left<\hat S_+\right>=S_+ =S\sin\beta e^{i\omega t+i\alpha} ~.
\label{precession}
\end{equation}
Here $\hat S_+$ is the operator of spin creation; $S_+=S_x+iS_y$; ${\bf S}=(S_x,S_y,S_z=S\cos\beta)$ is the vector of spin
density precessing in the applied magnetic  field ${\bf H}=H\hat{\bf z}$; $\beta$, $\omega$ and $\alpha$ are correspondingly the tipping angle, frequency and  the phase of precession. In the  modes under discussion, the magnitude of the precessing spin $S$ equals to an equilibrium value of spin density $S=\chi H/\gamma$ in the applied  field, where
$\chi$ is spin susceptibility of $^3$He-B, and $\gamma$
the gyromagnetic ratio of the $^3$He atom (the coherent state with half of magnetization $S=(1/2)\chi H/\gamma$ has been also observed in bulk $^3$He-B \cite{half}).
Similar to the conventional mass superfluidity which also experiences the off-diagonal long-range order, the spin precession in Eq.(\ref{precession}) can be rewritten in terms of the complex scalar order parameter \cite{FominLT19,V,B}
\begin{equation}
\left<\hat\Psi\right>=  \Psi=\sqrt{2S/\hbar}\sin \frac{\beta}{2}~e^{i\omega t+i\alpha}~,
 \label{OrderParameter}
 \end{equation}
If the spin-orbit interaction is small
and its contribution to the spectrum  of magnons  is neglected in the main approximation (as it typically occurs in $^3$He),
then $\hat\Psi$ coincides with the operator of the annihilation of magnons, with the  number density of magnons being equal to condensate density:
\begin{equation}
 n_M=\left<\hat\Psi^\dagger\hat\Psi\right>= \vert \Psi\vert^2 =\frac{S-S_z}{\hbar}~.
\label{NumberDensity}
\end{equation}
This implies that the precessing states in superfluid $^3$He realize the almost complete BEC of magnons. The small spin-orbit coupling  produces a weak interaction between magnons and leads to the interaction term in corresponding
Gross-Pitaevskii equation for the BEC of magnons (further we use units with $\hbar=1$):
  \begin{eqnarray}
 \frac{\delta F}{\delta \Psi^*}=0~,
  \label{GP}
  \\
 F=\int d^3r\left(\frac{\vert\nabla\Psi\vert^2}{2m_M} -\mu\vert\Psi\vert^2+{\bar E}_D(\vert\Psi\vert^2)\right),
\label{GL}
\end{eqnarray}
Here the role of the chemical potential $\mu=\omega-\omega_L$ is played by the shift of the precession frequency from  the Larmor value $\omega_L=\gamma H$; the latter may
slightly depend on coordinates if the field gradient is applied. In coherent states,  the  precession frequency $\omega$ is the same throughout the whole sample even in  the nonuniform field; it is determined by the number of magnons in BEC,  $N_M=\int d^3r n_M$, which is conserved quantity if  the dipole interaction is neglected. In the regime of continuous
NMR,   $\omega$ is the frequency of the applied RF field, $\omega=\omega_{\rm RF}$, and the chemical potential $\mu=\omega_{\rm RF}-\omega_L$ determines the magnon density. Finally, $m_M$
is the magnon mass; and ${\bar E}_D$ the  dipole interaction
averaged over the fast precession. The general form of ${\bar E}_D(\vert\Psi\vert^2)$  depends on the orientation of the orbital degrees of freedom described by the unit vector $\hat{\bf l}$ of the orbital momentum, see Ref.  \cite{BV}.

\section{Magnons BEC in bulk $^3$He-B and in non-deformed aerogel}

\begin{figure}
\begin{center}
\includegraphics[%
  width=0.65\linewidth,
  keepaspectratio]{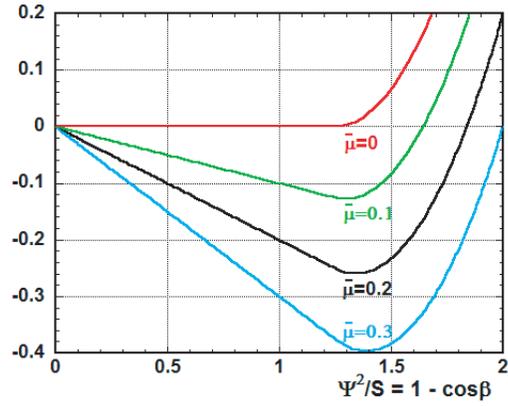}
\end{center}
\caption{(Color online) The Ginzburg-Landau    energy ${\bar E}_D(\vert\Psi\vert^2) -\mu\vert\Psi\vert^2$ in bulk $^3$He-B as a function of magnon density (tipping angle  of
precession) for different values of the normalized chemical potential $\tilde \mu= (\omega_{\rm RF}^2-\omega_L^2)/\Omega_L^2$,
Energy minima correspond to magnon BEC, i.e. coherent HPD states precessing with
frequency shift equal to the chemical potential} \label{DemoB}
\end{figure}

In the coherent precession in bulk $^3$He-B and in a non-deformed aerogel  \cite{DmitrievHPD}, the  spin-orbit coupling orients vector   $\hat{\bf l}$ along the axis of precession, i.e.  $\hat{\bf l}\parallel {\bf H}$. In this case the interaction term ${\bar E}_D$ expressed through the condensate order parameter has the form different from conventional 4-th order term in dilute gases \cite{V}:
 \begin{eqnarray}
 {\bar E}_D=0~,~\vert\Psi\vert^2<\frac{5}{4}  S~,
\label{0}
 \\
{\bar E}_D=\frac{8}{15}\chi\Omega_L^2 \left(\frac{\vert\Psi\vert^2}{S}-\frac{5}{4}
\right)^2~,~\vert\Psi\vert^2>\frac{5}{4}     S~.
     \label{FHPD}
  \end{eqnarray}
Here $\Omega_L\ll \omega_L$ is the Leggett frequency which characterizes the dipole interaction.
If the chemical potential $\mu$ is negative, i.e. $\omega$ is less than
$\omega_L$, the minimum of the Ginzburg-Landau (GL) energy ${\bar E}_D(\vert\Psi\vert^2) -\mu\vert\Psi\vert^2$ corresponds to $\Psi=0$, i.e. to the static  state
with non-precessing equilibrium magnetization ($\beta=0$).  For $\mu>0$ (i.e. for positive frequency shift), the profile of the (normalized) energy density is shown in Fig.~\ref{DemoB} for several values of $\mu$ given in dimensionless units $\tilde \mu =(\omega_{\rm RF}^2-\omega_L^2)/\Omega_L^2\approx \mu / (\Omega_L^2/2\omega_L)$.
The minimum of the GL energy corresponds to $\vert \Psi\vert^2/S = (5/4) + (15/32)\tilde \mu$. The consequence of the peculiar profile of the interaction term is that  as distinct from the dilute gases the formation of the magnon BEC starts with the finite magnitude $\vert\Psi\vert^2=(5/4)S$. This means that the coherent precession starts with a tipping angle  equal to the magic Leggett angle, $\beta=
104^\circ$, and then the tipping angle increases with increasing frequency shift.
 This coherent state  called the HPD   persists  indefinitely, if one applies a small RF field to compensate the losses of magnons caused by small spin-orbit interaction.

In conventional magnetic systems,  magnetization precesses in the   local field with the local frequency shift and thus experiences dephasing in the inhomogeneous field.
In the case of  magnon BEC,  the rigidity of the order
parameter (the gradient term in Eq.~\ref{GL})  plays an important role. The spatial dephasing leads to the gradient of chemical potential. This in turn excites the spin supercurrents, which finally  equilibrate the
chemical potential. In the steady state of magnon BEC the gradient of  a local field is
compensated by small gradient of magnon density $\vert \Psi \vert ^2$ in such a way, that the precession frequency  and its phase remain homogeneous throughout the whole sample.

\begin{figure}
\begin{center}
\includegraphics[%
  width=0.9\linewidth,
  keepaspectratio]{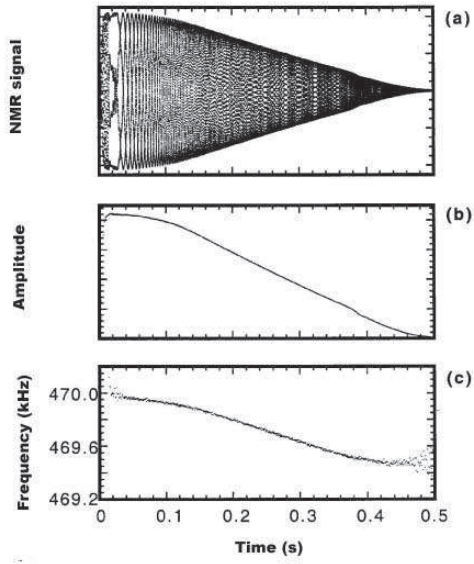}
\end{center}
\caption{(Color online) The typical signal of induction decay from
the BEC of magnons; (a) stroboscopic record of the signal; (b)
amplitude of signal; (c) frequency of the signal. Frequencies 469.95
and 469.4 kHz correspond to  Larmor frequency at the top and the bottom
of the cell. } \label{hpd2}
\end{figure}

In a pulsed NMR experiment,  the magnetization is deflected by a strong  RF pulse.
The typical induction signal after the pulse in the cell with  a large gradient of magnetic field along the axis of the cell is shown in Fig.~\ref{hpd2}. Due to the field gradient the induction signal should dephase and disappear in about 10 ms. Instead, after a transient process of about 2
ms, the induction signal acquires an amplitude corresponding to a 100\% coherent precession of the deflected magnetization with the spontaneously  emerging phase. This coherent state lives 500 times longer than the dephasing time caused by inhomogeneity.

\begin{figure}
\begin{center}
\includegraphics[%
  width=0.65\linewidth,
  keepaspectratio]{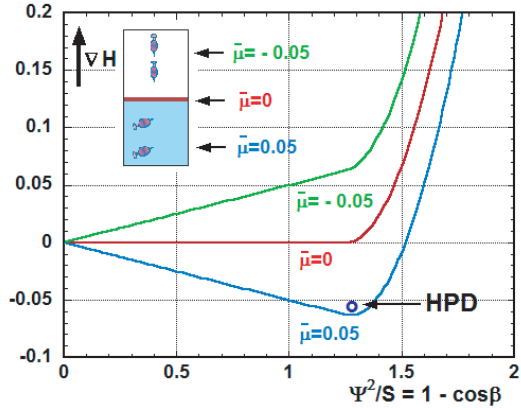}
\end{center}
\caption{(Color online) The  Ginzburg-Landau energy ${\bar E}_D(\vert\Psi\vert^2) -\mu\vert\Psi\vert^2$
as a function of magnon density for the same frequency $\omega$, but
at different Larmor frequency $\omega_L$ at the top and the bottom of the cell (this corresponds to different normalized chemical potential $\tilde \mu =(\omega^2-\omega_L^2)/\Omega_L^2$)} \label{HPD}
\end{figure}

What happens with the magnon system during the transient period? The
GL energy ${\bar E}_D(\vert\Psi\vert^2) -\mu\vert\Psi\vert^2$ in the presence of the gradient of magnetic field is shown in Fig.~\ref{HPD}. As soon, as at the higher field end of the
cell the deflection of magnetization becomes smaller than
104$^\circ$, the spin-orbit interaction cannot anymore compensate the
gradient of magnetic field. The cell splits into two domains, in one of them
the magnetization is stationary, while in other one magnons condense  with the density  close to $\vert \Psi\vert^2 = (5/4)S$, i.e. the magnetization precesses with $\beta$
slightly above  104$^\circ$.  In the subsequent process of relaxation caused by the non-conservation of magnon number,  the volume of the BEC condensate (HPD)  decreases. During the relaxation, the BEC  does not loose the phase coherence, but its chemical potential (precession
frequency) changes. The frequency of HPD corresponds to Larmor frequency
at the boundary of the domain (Fig.~\ref{HPD}) and slowly changes with relaxation as the boundary moves down (see Fig.~\ref{hpd2}). The amplitude of the signal exactly corresponds to the record of the frequency.

Of course, the HPD has been observed, studied and explained on the basis of theory of spin superfluidity and non-linear NMR long time ago
\cite{HPD}. However, the consideration of this phenomenon in terms of the
magnon BEC not only demonstrates
the real system with the BEC of excitations, but also allows us to simplify the problem and to study and search for the other types of the magnon BEC in $^3$He, such as $Q$-balls \cite{Qbal}; HPD$_2$ state found in a deformed aerogel \cite{HPD2}; coherent precession in $^3$He-A also found in the deformed aerogel \cite{JapCQS}; etc.

\section{Magnons BEC in superfluid $^3$He in aerogel}

In superfluid $^3$He-A, the spin-orbit interaction depends on the polar angle  $\beta_L$
 of the orbital vector $\hat{\bf l}$ in the following way:
\begin{eqnarray}
  {\bar E}_D= \frac {\chi\Omega_L^2}{4}
  \left[ -2\frac{\vert\Psi\vert^2}{S} +
  \frac{\vert\Psi\vert^4}{S^2}    +
    \left( -2+4 \frac{\vert\Psi\vert^2}{S}  -
  \frac {7}{4}\frac{\vert\Psi\vert^4}{S^2}\right)\sin^2{\beta_L}\right].
  \label{FDA}
  \end{eqnarray}
While the  sign of the quadratic term in Eq. ~(\ref{FDA})  is not
important because this term can be compensated by the shift of the chemical
potential $\mu$ in Eq.~\ref{GL}, the sign of the
quartic term is crucial for the stability of the BEC. In the static state (i.e. $\Psi = 0$) of a bulk
$^3$He-A,  the dipole energy is minimized when
$\hat{\bf l}$  is perpendicular to the magnetic field, $\sin\beta_L = 1$.
In NMR, this orientation of $\hat{\bf l}$ remains frozen, and the
quartic term remains negative. This attractive interaction between
magnons destabilizes the BEC, which means that homogeneous
precession of magnetization in $^3$He-A is unstable, as was
predicted by Fomin \cite{InstabAF} and observed experimentally
\cite{InstabAB}.

\begin{figure}
\begin{center}
\includegraphics[%
  width=0.65\linewidth,
  keepaspectratio]{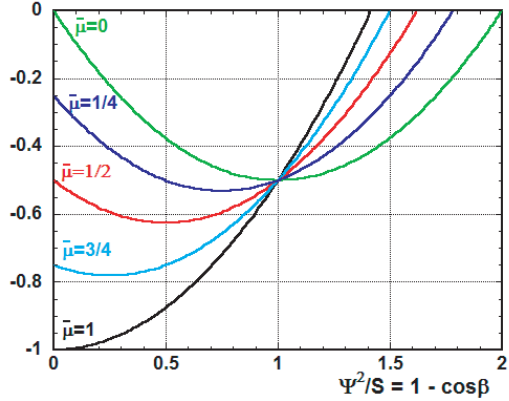}
\end{center}
\caption{(Color online) The normalized GL energy ${\bar E}_D(\vert\Psi\vert^2) -\mu\vert\Psi\vert^2$
 in superfluid $^3$He-A in the deformed aerogel, which orients the orbital
momentum $\hat{\bf l}$ parallel to the magnetic field. The GL energy is given as a function of magnon
density (tipping angle of precession) for different values of the normalized chemical potential $\tilde \mu =(\omega^2-\omega_L^2)/\Omega_L^2$. The minima of GL energy  correspond to the stable magnon BEC}
\label{CQSA}
\end{figure}

Recently, we  found the method, how to reorient the orbital
momentum $\hat{\bf l}$. For that we immersed $^3$He
inside the squeezed aerogel. Even a small deformation of a few
percent fixes the orbital momentum along the deformation.
\cite{JapA}. By applying the deformation along the magnetic field we
 fixed $\beta_L=0$, which made positive the quartic
term in Eq. ~(\ref{FDA}). As a results magnons can condense into
the stable  BEC state, as we observed experimentally
\cite{JapCQS}. For $\beta_L=0$, the GL energy ${\bar E}_D(\vert\Psi\vert^2) -\mu\vert\Psi\vert^2$
(Fig. ~\ref{CQSA}) is similar to the traditional GL energy for the conventional BEC in atomic gases.
This means that, as distinct from the  HPD in $^3$He-B, the minimum of the GL energy is realized even for small magnon densities, and in the experiments made in the
Institute of Solid State Physics in Kashiva, Japan, we observed
the formation of BEC in $^3$He-A without the domain formation.

In the 2007 experiments in Neel Institute, Grenoble we  applied
the aerogel deformation to reorient the orbital momentum $\hat{\bf l}$ in
$^3$He-B perpendicular to magnetic field. For this case the shape of the GL energy
allows for stable and unstable magnon BEC depending on the tipping angle $\beta$. Both  BEC states were observed in pulsed NMR experiments in a good agreement with the GL energy \cite{HPD2}.

\section{Magnon condensation  into Q-ball}

The non-topological soliton solutions, called $Q$-balls, have been
proposed by Coleman \cite{Col} as a mechanism of
formation of composite  particles in high energy physics. $Q$-balls appear  in theories
containing a complex scalar field with suitable self-interaction.
They are stabilized due to the conservation of the particle number,
or  in general the charge  $Q$. The $Q$-ball is a rather general
physical object (see \cite{Kus}), which in principle can be formed  in
different condensed matter systems. In particular, $Q$-balls were
suggested in the atomic Bose-Einstein condensates \cite{QBallBEC}.

\begin{figure}
\begin{center}
\includegraphics[%
  width=0.65\linewidth,
  keepaspectratio]{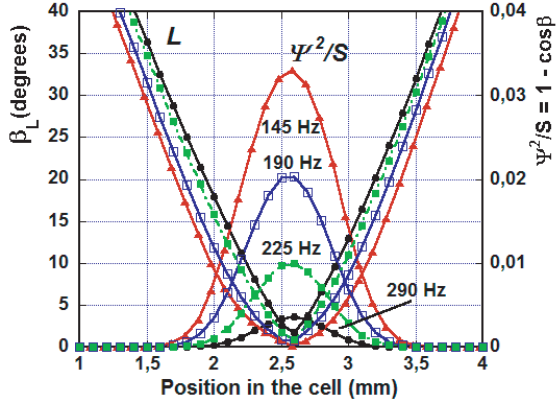}
\end{center}
\caption{(Color online) The angle of deflection of orbital momentum vector
$\hat{\bf l}$ (left scale) and  magnon density (right scale) as a
function of position in the one dimensional cell for different total
number of magnons $Q$. The NMR radiation frequency is shown for each
curve.} \label{quad}
\end{figure}

The $Q$-balls were found in NMR experiments in superfluid
$^3$He-B as a small-amplitude, but extremely long lived induction decay
signal \cite{PS}. The role of the
$Q$-charge is played by the total number of magnons $Q=\int d^3 r ~n_M$    \cite{Qbal}. In all previous
examples, magnons were almost homogeneously distributed inside the
BEC state. In $Q$-balls, the condensation of magnons occurs in a
local trap, produced by the texture of orbital momentum $\hat{\bf l}$. In the homogeneous state  with $\hat{\bf l}\parallel {\bf H}$ the spin-orbit interaction is zero for small $\Psi$ (see Eq. ~(\ref{0})), but in the $\hat{\bf l}$-texture $ {\bar E}_D\neq 0$:
\begin{equation}
  {\bar E}_D=\chi\Omega_L^2\left[   \frac {4\sin^2(\beta_L/2)}{5S}
\vert\Psi\vert^2- \frac {\sin^4(\beta_L/2)}{S^2} \vert\Psi\vert^4
\right],
     \label{FDQ}
  \end{equation}
where $\beta_L$ is  the polar angle of the vector $\hat{\bf l}$.
It contains the negative quartic term, which describes the attractive interaction between
magnons. This form does not support the Bose condensation in bulk
$^3$He, but supports formation of magnon liquid droplets. Such a  droplet may
form inside of the existing orbital texture trap. Also, the self-localized orbital trap can be spontaneously
created by magnons, in a full analogy with $Q$-balls. The
magnon liquid in the droplet and inside the $Q$-ball remains in the coherent precessing state.

The droplet or $Q$-ball is formed in places
of the sample where the potential
 \begin{equation}
  U({\bf r})=
\frac{4\Omega_L^2}{5\omega_L}\sin^2\frac{\beta_L({\bf r})}{2}~.
\label{Potential2}
\end{equation}
produced by the $\hat{\bf l}$-texture has a minimum.  The precise form of a $Q$-ball depends on the
particular texture and on the position in the container. Typically it is formed in the so-called flared-out texture realized in cylindrical  cells. Far from the
horizontal walls and close to the axis of the cell the angle
$\beta_L$ linearly depends on the distance $r$ from the axis:
$\beta_L({\bf r})\approx \kappa r$  (see review
\cite{SalomaaVolovik}). The potential (\ref{Potential2})
for magnons is $U(r)\propto \kappa^2r^2$, similar to the
harmonic trap used for confinement of dilute Bose gases.  The peculiarity of the attractive quartic term in the GL functional in Eq.~(\ref{FDQ}) is that it is not a constant but is $\propto
\kappa^4r^4$. This stabilizes the condensate in the trap: at fixed number $Q$ of magnons  the GL energy has minimum at some  $Q$-dependent size of the magnon droplet, $r(Q)$. This is the main feature of $Q$-balls. Fig. ~\ref{quad}  shows numerical simulations of the spatial distribution of magnon density inside the $\hat{\bf l}$ texture for
different $Q$ in one dimensional case. Note another feature of $Q$-balls: with increasing number $Q$ of the trapped magnons  the  potential well itself becomes modified by magnons and finally it is overwhelmingly  determined by $Q$.

Simultanious formation of many different
$Q$-balls has been observed in CW NMR \cite{Gren1}. They are
generated by spin waves in the broad range of frequencies. They can
be created in particular by topological defects at the walls of the
cell. As distinct from CW NMR, where the $Q$-balls are generated
starting continuously from $Q=0$, in pulsed NMR a $Q$-ball is formed
after a large $Q$ is pumped into the cell. In this case the 3D
$Q$-ball is often formed on the axis of the flared-out texture, away
from the horizontal walls \cite{Lan3}. Finally the lowest energy
$Q$-ball collects all the magnons and radiates as a single
harmonic oscillator with a slowly increasing frequency, due to
the modification  of the trap potential in the process of decay of the $Q$ charge, as shown in Fig.
~\ref{quad}. The relaxation of lowest energy $Q$-balls can be
compensated by additional RF pumping of magnons at any frequency above
the frequency of $Q$-ball, as was found in \cite{PS2} and explained
in \cite{Gren3}.

\section{Conclusion}

The variety of observed BEC states  of coherent precession in superfluid  $^3$He is  provided by   specific interaction terms in the GL energy, which in general are different from the conventional  4-th order term in the  traditional BEC. They come from the spin-orbit interaction and are governed by  the orbital vector $\hat{\bf l}$. This allows us to regulate the profile of the GL energy: by applied counterflow  \cite{DM2}; by deformation of aerogel \cite{JapCQS,HPD2}; by exciting the precession at one half of equilibrium magnetization \cite{half}; by using flared-out $\hat{\bf l}$-texture  \cite{Qbal} and topological defects; etc.

\begin{acknowledgements}

We are grateful to H. Godfrin and M. Krusius  for illuminating
discussions. This work was done as the result of collaboration in
the framework the Large Scale Installation Program ULTI of the
European Union (contract number: RITA-CT-2003-505313); the project
ULTIMA of ``Agence Nationale de la Recherche", France
(NTO5-2\_41909); collaboration between CNRS and Russian Academy of
Science (project 19058); and was supported in part by Russian
Foundation for Basic Research (grant 06-02-16002-a).
\end{acknowledgements}


\end{document}